\documentclass[journal]{IEEEtran}
\usepackage{graphicx}
\usepackage{array}
\usepackage[font={small}]{caption}
\begin{document}
\title{THE-FAME: \underline{TH}reshold based \underline{E}nergy-efficient \underline{FA}tigue \underline{ME}asurment for Wireless Body Area Sensor Networks using Multiple Sinks}
\author{S. Akram$^{1}$, N. Javaid$^{2,3}$, A. Tauqir$^{1}$, A. Rao$^{2}$, S. N. Mohammad$^{2}$\\\vspace{0.4cm}
$^{1}$Institute of Space Technology, Islamabad, Pakistan.\\
$^{2}$Dept. of Electrical Engineering, COMSATS Institute of IT, Islamabad, Pakistan.\\
$^{3}$CAST, COMSATS Institute of IT, Islamabad, Pakistan.}
\maketitle
\begin{abstract}
Wireless Body Area Sensor Network (WBASN) is a technology employed mainly for patient health monitoring. New research is being done to take the technology to the next level i.e. player's fatigue monitoring in sports. Muscle fatigue is the main cause of player's performance degradation. This type of fatigue can be measured by sensing the accumulation of lactic acid in muscles. Excess of lactic acid makes muscles feel lethargic.
Keeping this in mind we propose a protocol \underline{TH}reshold based \underline{E}nergy-efficient \underline{FA}tigue \underline{ME}asurement (THE-FAME) for soccer players using WBASN. In THE-FAME protocol, a composite parameter  has been used that consists of a threshold parameter for lactic acid accumulation and a parameter for measuring distance covered by a particular player. When any parameters's value in this composite parameter shows an increase beyond threshold, the players is declared to be in a fatigue state. The size of battery and sensor should be very small for the sake of players' best performance. These sensor nodes, implanted inside player's body, are made energy efficient by using multiple sinks instead of a single sink. Matlab simulation results show the effectiveness of THE-FAME.
\end{abstract}
\section{Introduction}
\IEEEPARstart{W}{BASN} started developing since 1995 to make communication at and near a human body. The whole idea behind developing these networks is to provide better health care services to critical patients and elderly people. With modern medicine and better lifestyle, the average expectation of life is increased. Now, society has more number of elder people, who need continuous monitoring and urgent medical aid. WBASN is excellent for this purpose with no or minimum intervention of humans.

Besides the health monitoring of elder people and critical patients with life threatening issues, WBASN can also be used in many other fields where continuous and remote monitoring of the health is necessary. These fields include players in different sports, astronauts in space, pilots, etc.

For health monitoring of athletes and other players, parameters like respiratory rate, heart rate, blood oxygen, blood glucose, body temperature, etc. are important. Another important issue with players is the fatigue of muscles. Commercially many sensors are available for fatigue measurements in humans like temperature and heartbeat measuring sensors. However, the sensors measuring amount of lactic acid in muscles can provide the most accurate results. This is because, when a muscle feel lethargic it requires more oxygen and this requirement is fulfilled by producing lactic acid as a by product.

The idea of accurately measuring the fatigue level in players can totally revolutionize the sports industry across the globe. The sports that include more physical exertion and include a team of players of different stamina, can especially be benefited. One of the most popular games that fall in this category is soccer. Fans and admirers get very upset when their favorite players show bad performance because of fatigue.

The sensors used for soccer players are in-vivo and also they should be of very small size for the ease of a player to show the performance. As the sensor has a very small size and low battery power, so, the transmission rate of data is limited. Generation and transmission of redundant data is minimized by using different threshold levels for sensors. Data packet is only generated by the sensor node when the sensed value is equal to or above the provided threshold level.

In our proposed protocol THE-FAME, we use two threshold values for players fatigue measurement. For players fatigue measurement, we use a threshold for lactic acid level in the blood and a threshold for total distance travelled by the player. In THE-FAME, sensors send data when they sense fatigue. This data is then routed towards the nearest sink present at the boundary of the ground.

Most commonly used wireless technologies in WBASN are ZigBee and bluetooth. Zigbee was developed to address the need for small sized nodes, low energy consumption and low cost wireless networks. It has the IEEE standard of 802.15.4. It operates in the unlicensed bands of 868 MHz, 915 MHZ and 2.4 GHZ.

Our work begins by defining the work related  with our technique in section II. Section III includes the motivation for the proposed scheme. In section IV, we explain basic fatigue measuring parameter and mention the speed profile of a soccer player during the match. Next, in section V, we show our proposed technique which limits the energy consumption of sensor nodes. We then present the simulation results of our technique in section VI, which depicts that it has minimum propagation delay and energy consumption. Finally, we conclude our work in section VII.
\section{Related Work}
The field of WBAN is not quite latest and only little research work is done in it. Those few research articles are mainly focused on patient's health monitoring. On the other side, if research for WBAN employed for player's fatigue monitoring has been checked, very few work can be found. Some of the articles on WBAN are listed below.

In \cite {kifayat2010body}, Kashif Kifayat $et$ $al.$ present a design and implementation of a body area sensor network and a gaming platform to dynamically adapt the physiotherapy treatments. The proposed framework use three components side by side which includes a Wireless Sensor Network (WSN), a game and a data acquisition manager. WSN collect information from the patients and forward it to the data acquisition manager which then provides updates to the game. As, time passes and the patient shows better mobility, the game reaches its higher levels and become difficult. This is done to train patients accurately with the passage of time.

M-ATTEMPT \cite{javaid2012m}, show an energy efficient routing protocol for heterogeneous WBASN attached on a human body for continuous patient monitoring. Sink is placed on the center of the body. M-ATTEMPT uses single hop data transmission for critical data and for normal data delivery to the sink, it uses multi-hop communication. It is a thermal aware protocol which when senses a hot-spot on the route, immediately change its route by skipping that node until it is back to its normal temperature. This will protect human tissues from being damaged.

Ashay Dhamdhere $et$ $al.$ provide real time team monitoring for a soccer match. \cite{dhamdhere2010experiments} present different challenges for monitoring soccer players. They give their own design where, they divide team members into different categories on the basis of their position from sink and then calculate delay and resource consumption profile of Wireless Body Area Network (WBAN) for each team member, respectively.

L. De Nardis $et$ $al.$ in \cite{de2010mobility} use body area networks for soccer players to generate a realistic mobility model. Each player has Body Area Network (BAN), which provides player's position information throughout the game. The communication is done using inter-BAN multi-hoping. They designed a new mobility model called DynaMo and compared its delay and throughput with Reference Point Group Mobility model (RPGM).

In \cite{sampangi2011novel}, authors work on WBAN and the factors affecting the reliability of data. According to the paper, there are two main factors that affects the data reliability namely, data freshness and data accuracy. Authors employed multiple sinks on a single human body to maintain freshness of data for remote monitoring of critical patients. The associated delays and packet loss also decrease because of their proposed scheme.

Yeongjoon Gil $et$ $al.$ introduce a multi-body sensor platform for tele-medicine and emergency scenarios in \cite{gil2012synchronous}. They present a prototype for their work. The results show that monitoring only a single parameter for a patient is not sufficient because, mostly diseases and ailments are interconnected. So, a framework for measuring multiple signals from a body is helpful specially when it is low power consuming and not harmful for any human tissue.

Authors in \cite{rahim2012adaptive}, \cite{javaid2013energy},\cite{ain2012modeling}, \cite{manzoor2012noise} and \cite{javaid2012performance} work on MAC layer of the OSI model and give a medium access control protocol for WBAN, give a through survey on energy efficient protocols for WBAN, present arm mobility for patients in WBAN, noise filtering for WBAN channel and provide localization technique for WBAN, respectively.

N.S.A. Zulkifli $et$ $al.$ implement heart rate monitoring algorithm for players, performing strenuous exercises \cite{zulkifli2012xbee}. Heart rate monitors (HRM) are commonly used as a training aid for sports persons. When HRMs are incorporated with WBAN, they can provide continuous and accurate monitoring of players during hard exercises.

Authors in \cite{baca2010server} introduce a prototype system for remote monitoring of athlete's performance by providing live feedback. Each player is equipped with sensors and a mobile device. Some specific parameters are measured and then buffered locally. After a certain time period, the data will be forwarded using internet to the coaches.

Samuel J. Lerer $et$ $al.$ provide a personnel management system for collecting vital health information in ice-hockey players \cite{lerer2010building}. Ice hockey is a very strenuous and a difficult game because of its requirements for high speed. The paper focuses on using inexpensive sensors for college students playing ice hockey. It uses a respiratory rate monitoring system and use audio processing on the collected data to find fatigue in players.

Miguel Garcia $et$ $al.$ in \cite{garcia2011wireless} propose a new remote monitoring mechanism for soccer players, called Wireless Soccer Team Monitoring (WSTM). They use multi-hoping topology for data routing in the network. Two sinks are placed behind the goals and each player is equipped with a BAN to calculate fatigue.
\section{Motivation}
In any sport, one of the major concerns is the fitness of a player. Whereas, in team sports like soccer or hockey, fitness of each player is not only important but it is also considered a match winning criteria. The problem with these sports is that the fatigue level of a player is continuously changing during the play as it includes running, sprinting and many other tiring activities. So, there a mechanism is needed to constantly monitor each player in the team. Whenever any critical condition occurs, coaches and health officials can immediately take precautionary measures to handle the situation and reduce the chance of any further injury.

In \cite{garcia2011wireless} a protocol for soccer players monitoring is introduced which uses multi-hop routing to transmit data to the Base Station (BS). However, large delay occurs when the player is not present near the BS.
\cite{lerer2010building} discuss player monitoring in ice hockey based on audio processing of respiratory data of players. However, monitoring respiratory data needs the use of microphones which creates difficulties for players. Also, its processing creates long delays and produces less accurate results.
Keeping this in mind, we propose a new protocol called THE-FAME. Our proposed protocol produces less propagation delays and increases the lifetime of the entire network.
\section{Experimental Soccer Match Details}
The standard size of a soccer ground can vary from 100 yards to 120 yards in length and from 60 yards to 80 yards in width. We choose a size of $106 \times 68$ square yards for the soccer field. The total number of players from both teams are 22 i.e. 11 players per team.
\begin{figure}
  \centering
  \includegraphics[scale=0.4]{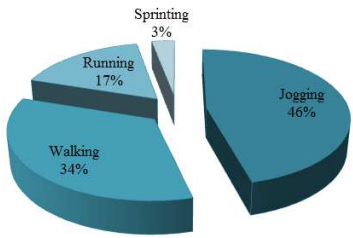}\\
  \caption{Soccer player's speed profile.}\label{1}
\end{figure}
There are many game strategies but the most common are defensive and attacking. Defensive game has 4 to 5 players in their own half and rest of the players are distributed according to the need. Whereas, in attacking situation 4 players play in the other team's half to score more goals.
The speed and fatigue analysis of players in the field is discussed in the following two subsections.
\subsection{Soccer Player Speed Analysis}
A soccer match on average continues from 90 to 100 minutes. During this time, the total distance covered by a professional player is slightly more than 11 km per match (nearly 7 miles). This distance is not only covered by running but, some sprinting is also performed \cite {url5} as shown in the statistics provided in Fig. 1. The average running speed of a player can vary from 10.3 km/hour. to 12.9 km/hour. During the possession of a ball, soccer players usually sprint and can achieve a speed of 25 km/hour. \cite{url1}

A soccer player attempts approximately 100 sprints per match lasting about 2-5 seconds. On average, for a single match, the minimum work to rest ratio for a soccer player is 1:2 \cite{url2}.
Soccer player's speed, direction and ball possession can all be compromised if he cannot recover from the fatigue. If a player continues running instead of taking rest then, he can be seriously injured which can lead to permanent damage of muscle's tissues. So, to avoid this, a threshold for fatigue level must be defined beyond which, players can face serious health issues.
\begin{figure}
  \centering
  \includegraphics[scale=0.5]{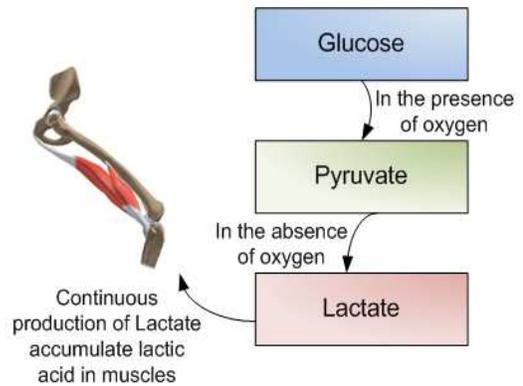}\\
  \caption{Lactic acid production process in muscles}\label{2}
\end{figure}
\subsection{Muscle Fatigue Measurement}
Muscle fatigue is commonly explained as the inability of muscles to generate any force. When humans perform any vigorous exercise or do sprinting or swimming, they begin to inhale faster to send more oxygen to the working muscles. A human body generates most of the energy using aerobic methods. But under some circumstances where, the amount of oxygen ($O_{2}$) required is more than the need, the body uses anaerobic methods i.e. it uses glucose by a process called glycolysis to produce energy. In the process of glycolysis, glucose is metabolized into a substance called pyruvate. In presence of enough $O_{2}$, pyruvate is further broken down to generate energy, but when there is a limited supply of $O_{2}$ in the body, pyruvate changes into lactate which will then be transformed into energy. The process is also explained in Fig. 2.

If cells in the muscles do the above mentioned practice at high rates, i.e. for more than 3 minutes, lactic acid will start to accumulate into muscles. This shows that it is a good parameter to measure fatigue in a soccer player \cite{url3}.
The sensor used to measure lactic acid level, is an in-vivo sensor which includes a needle, pricking the muscle to draw blood. Normal results show a value between 4.5 to 19.8 mg/dL (0.5-2.2 mmol/L).
where, mg/dL = milligrams per deciliter and mmol/L = millimoles per liter \cite{url4}.

\section{Proposed Technique}
THE-FAME works on the network layer of the standard OSI model. We take two teams of 12 players each, for the experimentation. Each player has an implanted sensor to collect his lactic acid level regularly as shown in Fig. 3. In THE-FAME, each player's fatigue information is transmitted whenever, the maximum threshold level for lactic acid in blood or distance covered during the match, reaches. This information is routed towards the sink using only a single hop. Single hop or direct transmission will be useful for decreasing the propagation delay of the network.
\begin{figure}
  \centering
  \includegraphics[scale=0.4]{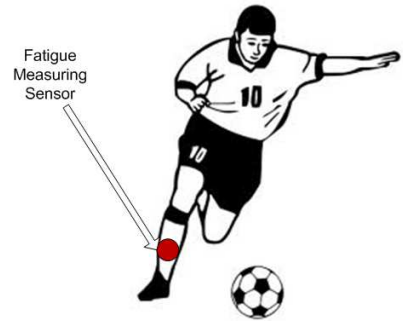}\\
  \caption{Soccer player.}\label{3}
\end{figure}
To support direct transmission mechanism, we use 6 sinks fixed at the boundaries of the ground whose locations are given in Table I. Multiple sinks in the field let the sensors consume less power during transmission of fatigue information.
\begin{table}
\centering
\begin{tabular}{|c|c|c|}
  \hline
  $Sink$&\multicolumn{2}{c}{$Axis$}
  \vline\\
  \cline{2-3}
  $Number$&$X$ $(yards)$&$Y$ $(yards)$\\
  \hline
  1&0&34\\\hline
  2&17&0\\\hline
  3&51&0\\\hline
  4&106&34\\\hline
  5&17&106\\\hline
  6&51&106\\\hline
\end{tabular}
\caption{Location of sinks at the soccer ground}
\end{table}

In THE-FAME, we use RPGM model to show the movement of players in the ground. It is a commonly used mobility model for soccer players. It is given a maximum speed of 25 km/hour for players mobility and a normal running speed of 10-12 km/hour.
\begin{figure}
  \centering
  \includegraphics[scale=0.4]{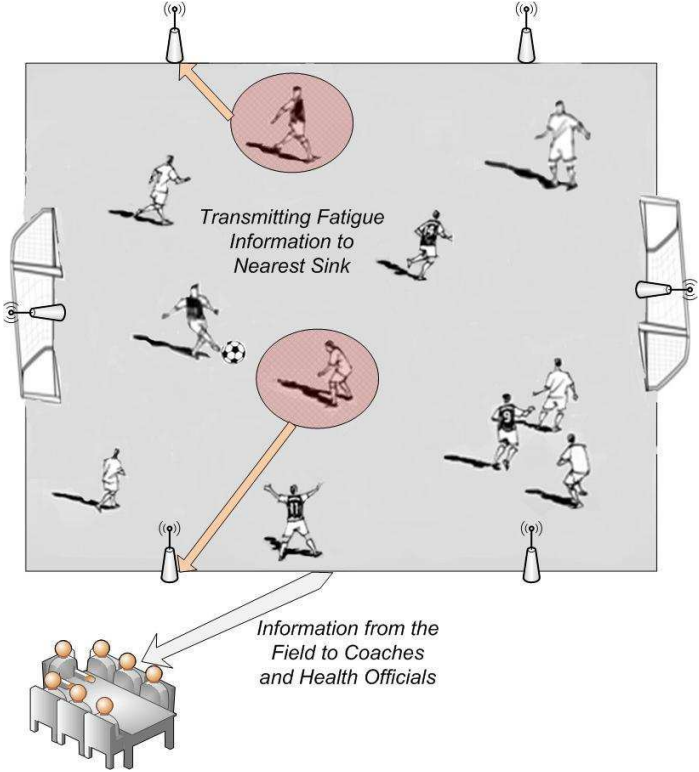}\\
  \centering
  \caption{Fatigue measuring process during the game.}\label{4}
\end{figure}
In THE-FAME, we use direct transmission method as the network topology. This method is supported by using multiple sinks at the border of the ground. To show a comparison between energy requirements of direct and multi-hop communication, the equations for both transmission methods are stated below:
\begin{equation}\label{1}
  E_{MH_{transmit}}(k,d) = N \times (E_{circuitry}+E_{amp})\times k \times d^2
    \end{equation}
  \begin{equation}\label{2}
  E_{MH_{receive}} (k) = (N-1) \times (E_{circuitry}+E_{amp}) \times k
\end{equation}
\begin{equation}\label{3}
  E_{MH_{Total}} = E_{MH_{transmit}}+E_{MH_{receive}}
\end{equation}
\begin{equation}\label{4}
  E_{DT_{transmit}}(k,d) = (E_{circuitry}+E_{amp}) \times k \times d^2
    \end{equation}
\begin{equation}\label{5}
  E_{DT_{Total}} = E_{DT_{transmit}}
\end{equation}
where, $E_{circuitry}$ is the electrical energy consumed by the circuit, $k$ is the packet size and $d$ is the distance from sensor node to sink. $N$ in the first three equations, represents the number of hops needed to reach sink. Equations (4) and (5) clearly show that the total energy is N times less for a direct transmission scenario as compared to multi-hopping, given in equations (1), (2) and (3). Here, one can argue about the distance factor present in all the equations, as, these distances are very small for multi-hopping. But this problem can be catered for by using multiple sinks along the field. Use of multiple sinks decreases the distance from sensor to sink.
The whole process of measuring fatigue during a live soccer match and the deployment of sinks at the boundary of the ground is well depicted through Fig. 4.
\begin{figure}
  \centering
  \includegraphics[scale=0.2]{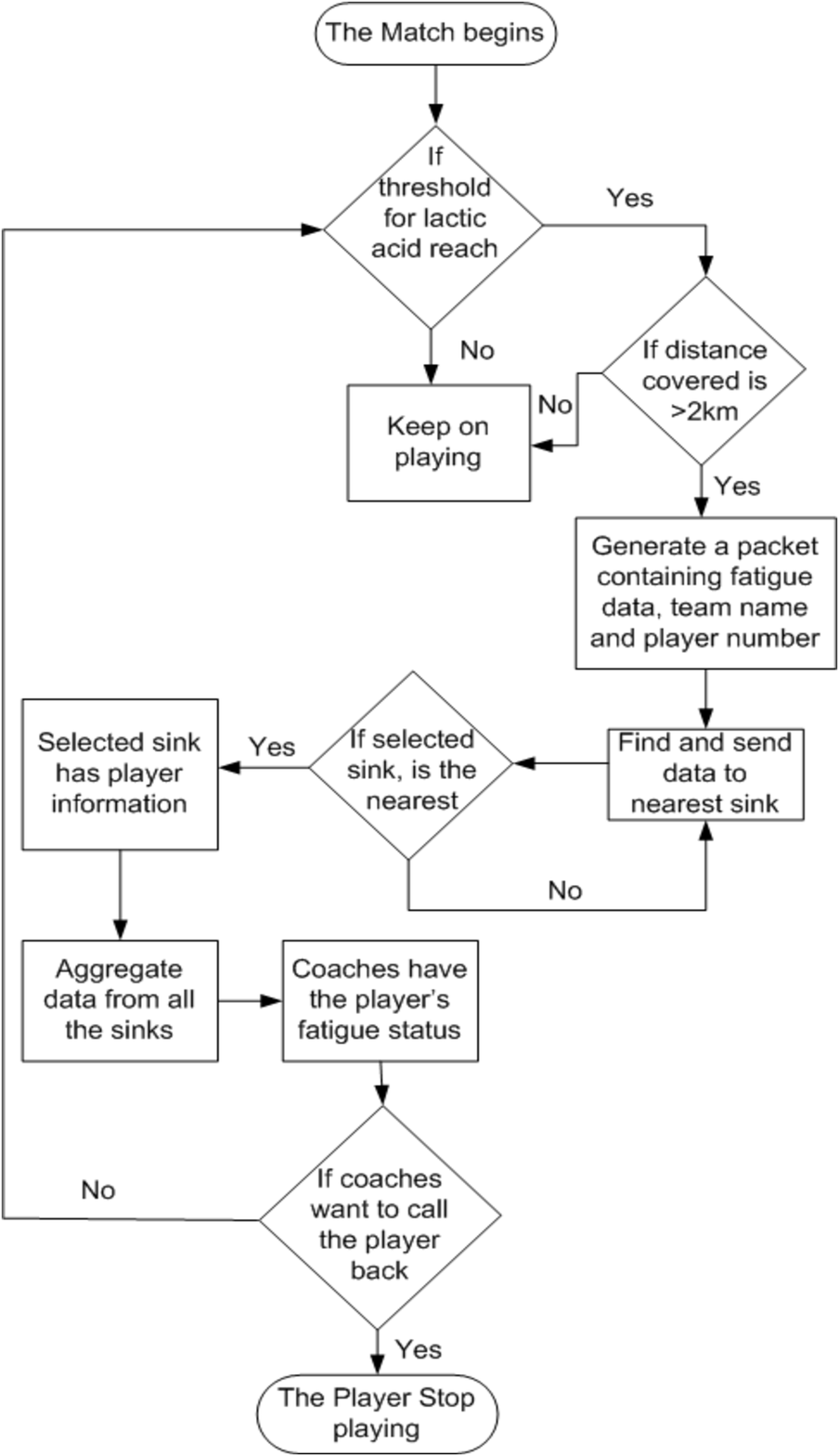}\\
  \caption{Flowcart of the proposed technique}\label{5}
\end{figure}

\section{Simulation Results}
The simulations presented in this paper are performed using MATLAB. We have created a soccer field of $106 \times 68$ square yards. Players mobility is shown using RPGM model. The positions of sinks are given in Table I. We choose a packet size of 1024 bits for transmitting important informational packets to sinks.
After sending data to the nearest sink of each player, all the data from these sinks is aggregated at the sink data aggregation unit and then sent to the monitor used by the team's coach and health experts. The flow chart of the process of finding fatigue in a soccer player and the whole network topology is explained in Fig. 5.

The differences in the network topologies and some parameters that are used in both protocols are stated below and also listed in Table II:
\begin{itemize}
  \item  WSN Soccer Team Monitoring protocol (WSTM) \cite{garcia2011wireless} uses multi-hop communication for players data delivery to the BS. Literature shows that the reception of packets consume nearly as much energy as in transmission which doubles the burden on the sensor nodes and hence, they die quickly in multi-hop scenario. Whereas, in THE-FAME, we use a single-hop technique.
  \item  Another difference between THE-FAME and WSTM, is that we use a composite parameter which includes a threshold based mechanism to transmit data to the nearest sink for a particular sensor node. Whenever the threshold for fatigue in a particular player or covered distance reaches, the sensor senses it and immediately transmit information to the concerned authorities. On the contrary, WSTM uses a period of 10 seconds after which it transmits data to the BS on a regular bases.
  \item  WSTM uses only two sinks present behind each goal. THE-FAME on the other hand, has six different sinks placed at the boundaries of the soccer ground. The location of the sinks is carefully chosen in order to have maximum throughput of the network and minimum propagation delay for the transmitted data packets.
  \item  The parameter we use in our protocol to find fatigue in a player's muscle is the amount of lactic acid accumulation in his/her muscle. WSTM doesn't mentioned any parameter on which they based their whole protocol.
  \item THE-FAME uses a frequency of 13.56 MHZ for data transmission because it is a lower range frequency and is not hazardous for human tissues. Whereas, WSTM uses a frequency of 2.4 GHz.
\end{itemize}

\begin{table}[h]
\centering
\begin{tabular}{c|c}
  \hline\hline
  \bf{THE-FAME} & \bf{WSTM}\\
  \hline
  Operating frequency & Operating frequency\\
  of sensors = 13.56 MHZ &  of sensors = 2.4 GHZ \\\hline
  Number of sinks = 6 & Number of sinks = 2 \\\hline
  Sinks are scattered & Sinks are placed\\
  on border of ground & behind each goal\\\hline
  Direct transmission & Multi-hop transmission\\\hline
  Uses a composite parameter & Parameters for team\\
  for fatigue measurement & monitoring are not specified \\\hline
  Uses a threshold for & No\\
  transmission of fatigue messages &  threshold defined  \\\hline
  Transmission occurs on & Transmission occurs \\
  specified events & every 10 sec  \\\hline
  In-vivo sensor is used & sensors are not specified \\
  \hline\hline
\end{tabular}
\caption{Differences between THE-FAME and WSTM}
\end{table}

In THE-FAME, we use direct transmission with multiple sinks to avoid the drain of battery. As a result, as shown in Fig.  6, nodes live longer and need not to be changed frequently.
However, in WSTM sensor nodes die quickly as compared to our proposed protocol. The difference in results is because, WSTM uses multi-hop scheme to route its data towards the sink, whereas, our protocol uses direct transmission method with multiple sinks in the field.

Stability period is the amount of time elapsed before the first sensor node in the network die. THE-FAME shows that the first node die nearly at 5100th round, as compared to the WSTM protocol in which the first node is dying after 2700 rounds. These analysis show that the parameter known as the stability period, is very effective in order to clearly understand the true behaviour of any wireless network. Fig.6 clearly reveals that our proposed protocol is 22$\%$ more stable than WSTM.
\begin{figure}
  \centering
  \includegraphics[scale=0.6]{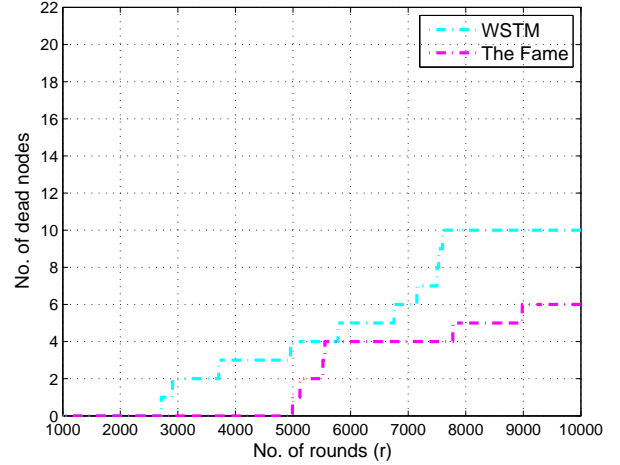}\\
  \caption{Number of dead nodes in the network.}\label{6}
\end{figure}
\begin{figure}
  \centering
  \includegraphics[scale=0.6]{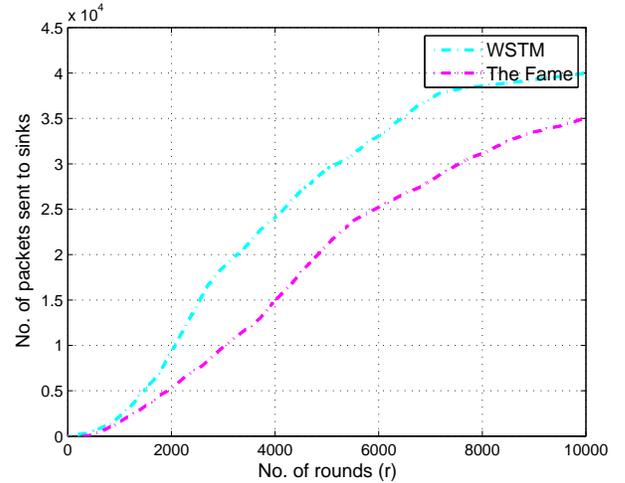}\\
  \caption{Total number of packets sent to sinks.}\label{7}
\end{figure}
\begin{figure}
  \centering
  \includegraphics[scale=0.6]{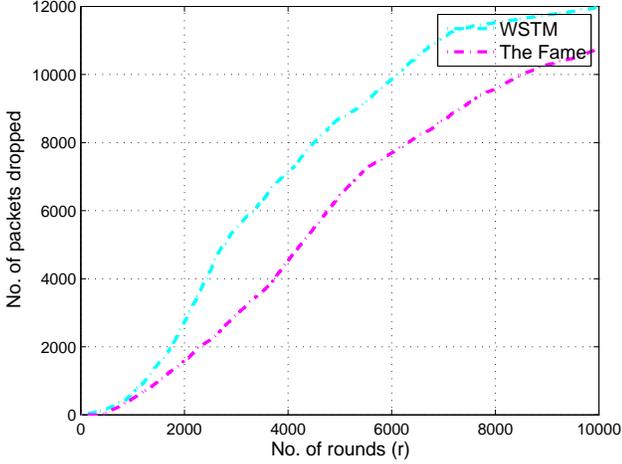}\\
  \caption{Packets dropped in the network.}\label{8}
\end{figure}

As explained earlier, in THE-FAME sensor nodes only transmit data to the sink when the threshold for fatigue is reached for the particular player.
In \cite{garcia2011wireless} nodes sends data directly to the BS if it is at its minimum distance to BS, like in the case of a goal keeper, otherwise it uses other players as relay nodes to route data towards the BS. On the other hand, THE-FAME only uses direct transmission method to send data to the BS. Hence, there are more packets generated in the network because of multi-hopping used in WSTM as given in Fig. 7.

In real world no channel is an ideal channel and especially in the case of wireless medium, there is always some loss of data in the network because of multi-path effects like refraction, reflection and absorbtion etc. Keeping this in mind we have chosen a noisy channel for simulation purpose instead of an ideal one, to make it realistic with a packet drop probability of approximately 30$\%$. Fig. 8 shows the amount of packets dropped in the network before reaching any sink.
Packets received successfully at the sinks are roughly about 70$\%$. The amount of packets received at all the sinks without any error is another important parameter to look for in the network. This parameter shows the accuracy of the protocol and network topology. This parameter for both protocols is presented in Fig. 9.

\begin{figure}
  \centering
  \includegraphics[scale=0.6]{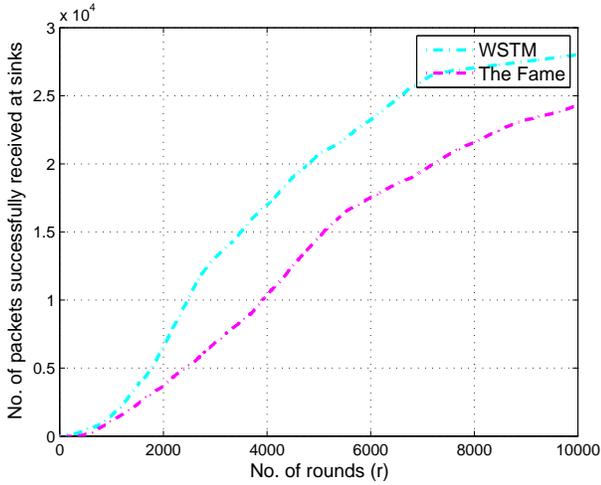}\\
  \caption{Packets received at sinks successfully.}\label{9}
\end{figure}
\begin{figure}
  \centering
  \includegraphics[scale=0.6]{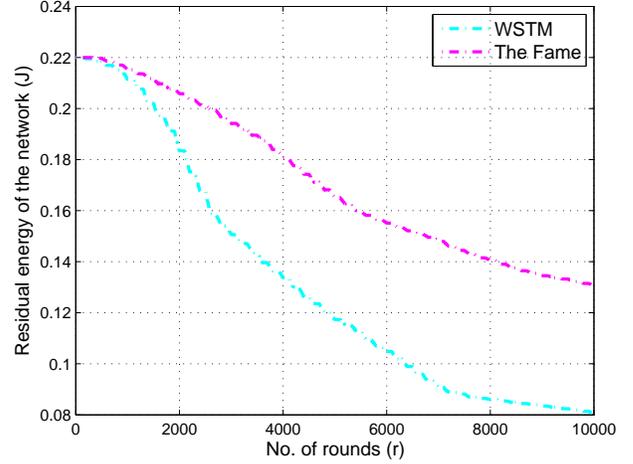}\\
  \caption{Residual energy of the network.}\label{10}
\end{figure}
Residual energy shown in Fig. 10 depends on the overall life time of the sensors in the network. In WSTM as, the nodes consume more energy during the communication process because of multi-hop scheme, so the residual energy deplete very quickly. Whereas, in THE-FAME protocol, the nodes live longer despite of the fact that the nodes have very low initial energy in our proposed scheme.
\begin{figure}
  \centering
  \includegraphics[scale=0.6]{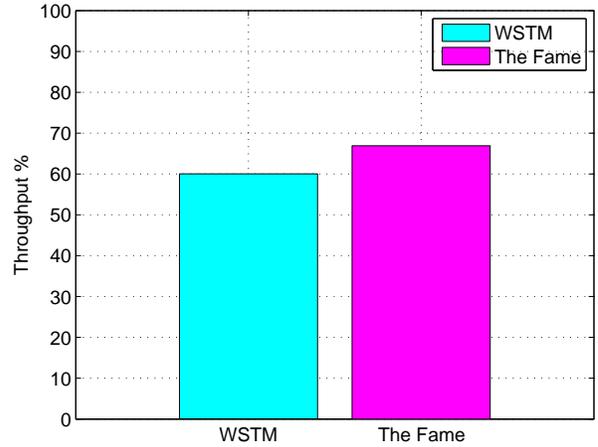}\\
  \caption{Throughput.}\label{11}
\end{figure}
\begin{figure}
  \centering
  \includegraphics[scale=0.6]{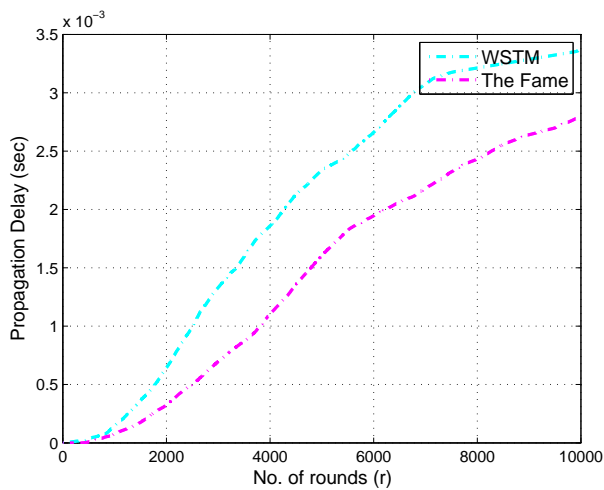}\\
  \caption{Propagation delay in the network.}\label{12}
\end{figure}

Fig. 11 compares the throughput of 2 schemes, THE-Fame and WSTM.
Throughput of a network is given by the following equation:
\begin{equation}\label{6}
\rm{Throughput}(\%)= \frac {\rm{Packets } \rm{Received}}{\rm{Total } \rm{Packets } \rm{Transmitted}} \times 100
\end{equation}

From Fig. 11 we are confident to say that the packet delivery ratio is 8 $\%$ better in THE-FAME as compared to WSTM.
Throughput of THE-FAME, reveals that our proposed protocol is more efficient.

It is very obvious from the Fig. 12 that THE-FAME has less propagation delay than WSTM, because it uses direct transmission method to send information to the sink unlike WSTM, which uses multi-hop and generates more delay. Propagation delay is an important factor to handle in scenarios where, we want to send critical data more quickly. THE-FAME only transmits data when, a threshold for muscle fatigue is reached.
In this respect, The-Fame should be preferred to achieve minimum propagation delays.

\section{Conclusion}
In this paper, we present an energy efficient fatigue measuring protocol for soccer players using WBASN. Direct transmission is selected to sent data to the BS to achieve minimum delay. The problem with direct transmission is more energy consumption, which is eradicated using multiple sinks along the border of the ground. We also use a composite parameter which consist of thresholds for lactic acid and distance covered by a player. The use of this composite parameter makes fatigue measurement more reliable. Hence, THE-FAME gives better results in both energy and delay profiles as compared to multi-hop routing scheme. In future work, we will implement Expected Transmission Count (ETX) link metrics as demonstrated in~\cite{javaid2009performance}~\cite{dridi2010performance}~\cite{dridi2009ieee}~\cite{dridi2009ieee1}.

\end{document}